# An Overview of Ontologies and Tool Support for COVID-19 Analytics

Aakash Ahmad
*College of Computer Science and Engineering*
*University of Ha'il*
Saudi Arabia
a.abbasi@uoh.edu.sa

Madhushi Bandara
*School of Computer Science*
*University of Technology Sydney*
Australia
Madhushi.bandara@uts.edu.au

Mahdi Fahmideh
*University of Southern Queensland*
Australia
Mahdi.Fahmideh@usq.edu.au

Henderik A. Proper
*Luxembourg Institute of Science and Technology*
and *University of Luxembourg*Luxembourg
e.proper@acm.org

Giancarlo Guizzardi
*Faculty of Computer Science*
*Free University of Bozen-Bolzano*
Italy
giancarlo.guizzardi@unibz.it

Jeffrey Soar
*University of Southern Queensland*
Australia
Jeffrey.Soar@usq.edu.au

*Abstract*—
**Context**: The outbreak of the SARS-CoV-2 pandemic of the new COVID-19 disease (COVID-19 for short) demands empowering existing medical, economic, and social emergency backend systems with data analytics capabilities. An impediment in taking advantages of data analytics in these systems is the lack of a unified framework or reference model. Ontologies are highlighted as a promising solution to bridge this gap by providing a formal representation of COVID- 19 concepts such as symptoms, infections rate, contact tracing, and drug modelling. Ontology-based solutions enable the integration of diverse data sources that leads to a better understanding of pandemic data, management of smart lockdowns by identifying pandemic hotspots, and knowledge-driven inference, reasoning, and recommendations to tackle surrounding issues.
**Objective**: This study aims to investigate COVID-19 related challenges that can benefit from ontology-based solutions, analyse available tool support, and identify emerging challenges that impact research and development of ontologies for COVID-19. Moreover, reference architecture models are presented to facilitate the design and development of innovative solutions that rely on ontology-based solutions and relevant tool support to address a multitude of challenges related to COVID-19.
**Method**: We followed the formal guidelines of systematic mapping studies and systematic reviews to identify a total of 56 solutions – published research on ontology models for COVID-19 – and qualitatively selected 10 of them for the review.
**Results**: Thematic analysis of the investigated solutions pinpoints five research themes including telehealth, health monitoring, disease modelling, data intelligence, and drug modelling. Each theme is supported by tool(s) enabling automation and user-decision support. Furthermore, we present four reference architectures that can address recurring challenges towards the development of the next generation of ontology-based solutions for COVID-19 analytics.

## I. INTRODUCTION

Throughout history, pandemics have ravaged humanity with plagues and infections that created humanitarian crises, severed social ties, hindered economic growth, and caused loss of human lives [1]. With the most recent outbreak of COVID-19 pandemic, states, communities, and individuals are facing war-like situation – with economic recession and an exponentially growing infection rate – that needs to be restrained [2]. Such an 'invisible enemy' has forced the world into a grand lockdown that has not been experienced in the past; as of now, COVID-19 has infected more than 207 million people with 4.35 million recorded deaths. Researchers and practitioners across various domains such as medical and life sciences, economics, and engineering are striving to put forward solutions to counter such a threat and aid the society in coping with the fallbacks [3]–[5].

In the same context, researchers and practitioners in knowledge management communities face a challenge about *how ontology-based systems can be exploited to tackle the current pandemic?* Ontology-based systems utilise semantic models, data storage and processing software as well as user-oriented analytics tools to enable information integration, multi-modal data analysis, and data visualisation related to COVID-19 infections [6]–[9].

They can help annotating real-time context-sensitive data from different sources such as tracing apps, performing context-aware data integration, and analytics functions that aid in taking the right actions to protect citizens and prevent transmission of the disease [10], [11]. The aim of this paper is to investigate existing ontology-based solutions proposed by the research community and evaluate their capabilities that assist in analysing and managing COVID-19 spread. The primary contributions of this research are to:

- Investigate the role of ontology-based solutions in software systems to address the challenges related to COVID-19 data.
- Evaluate the available tool support that automates and customises (i.e., enabling user decision support) to com-

plement ontology-based solutions.
- Present reference architectures as a generic solution to address the recurring problems for COVID-19 analytics.

To our knowledge, there is no existing systematic investigation, classification, and comparison of ontology-based solutions and tools support for COVID-19 analytics. Furthermore, the presented reference architecture models can provide guidance to develop software systems exploiting ontologies to address the issues related to COVID-19 pandemic.

The rest of the paper is organised as follows. Section II presents the context and background details of this study. Section III presents the methodology. Section IV presents the results of this study. Section V discusses the key findings of this research. Section VI concludes the paper.

## II. RELATED WORK

With a widespread proliferation of COVID-19 infections and their socio-economic impacts across the globe, nations and their administrative stakeholders are striving hard to exploit existing disaster management infrastructures [3] or smart city frameworks [4] to counter the pandemic. Moreover, novel approaches such as those that unify big data and artificially intelligent systems [5] accumulate an unprecedented amount of data from public health monitoring to support data-driven intelligence for pandemic management. We classify the most relevant existing work as (i) modelling of pandemic data (Section II. A), (ii) ontological management of pandemic (Section II. B), and (iii) analytics for COVID-19 (Section II. C).

### A. Modeling Pandemic Data

Some recent research and development efforts [1] represent a catalogue of solutions for data-driven modelling of pandemic scenarios along with decision strategies to counter the impacts of the pandemic on health care, social norms, and economic downturns. Modelling pandemic data (e.g., infection tracing, spread rates, and social distancing, etc.) is fundamental to visualise, simulate, and analyse pandemic spread as demonstrated by several studies conducted across the globe from Asia [2] to Europe [12], Americas [6], and Africa [13]. Specifically, Kaxiras & Neofotistos [7] and Bastos & Cajueiro [6] indicate how mathematical models helped simulate infection growth and contact tracing in countries like India and Brazil that represented the epicenter of the pandemics. For example, Bastos & Cajueiro [6] modelled data collected from public health and surveillance units to simulate long-term scenarios of the pandemics that reflect different levels of engagement of the Brazilian social distancing policy. Multi-disciplinary research efforts including but not limited to mathematical models, software tools, biological structures, and chemical compositions have been synergised for infection modelling, and spread prediction to the simulation of contact tracing [1], [7]. Despite the strategic benefits of pandemic modelling, several issues such as performance, scalability, and accuracy represent the potential limitations of model-based solutions [1], [2], [6], [13].

### B. Ontological Reasoning for COVID-19 Data

Several ontology-based techniques have been proposed to conceptualise fundamental concepts of COVID-19 data (e.g., infection symptoms or patients' health) and define essential relationships between these concepts to enable automated reasoning about data [10], [14]. The Infectious Disease Ontology (IDO) is a suite of interoperable ontology modules that aims to provide coverage of all aspects of the infectious disease domain, including biomedical research, clinical care, and public health. IDO provides foundations for ontology-based reasoning for COVID-19 use-cases and supports reproducibility of infectious disease research [14]. Some existing solutions [10], [15] have extended IDO – enabling data representation and automated reasoning – to support safety monitoring for indoor individuals and to analyse patients' data about the COVID-19 pandemic. Specifically, COVID-19 Ontology for cases and patient information (CODO) extends IDO to enable the representation of patient and infection data to create a network that supports the behaviour analysis of the disease, possible paths of disease spread, and various factors of disease transmission [10].

Such ontology-based models are useful because they enable (i) structural representation of COVID-19 data, (ii) semantics of the data, and (ii) behavioural analysis of data to support predictive analytics for pandemic spread [11]. The ontologies in [11], [15] leverages big data analytics techniques to extract significant information from multiple data sources for generating real-time statistics for contactless temperature sensing, mask detection, and monitoring social distancing.

### C. COVID-19 Analytics

The research on addressing technical challenges associated with COVID-19 analytics is still in its early phase and primarily focused on survey-based studies investigating the applications of existing solutions [8], [16] or proposing data analytics methods for sensing, mining, and visualising data from healthcare units [17], [18]. Most recently, a multitude of survey-based studies have been published on investigating contact tracing apps [8], COVID-19 dashboards for data visualisation [19], and predictive analytics for infection spread and social distancing [16]. Such studies streamlined needed solutions and provided recommendations to develop solutions for predictive analytics of COVID-19 [16]. A few works have relied on big data analytics for real-time contact tracing based on travel history and clinical symptoms of potential infections [17], [19]. For example, to support real-time visualisation of infection spread, C2SMART team exploited data mining and cloud computing techniques to investigate the impact of COVID-19 on mobility and sociability based on New York City and Seattle based case study [19]. The solution provides a dashboard for interactive data analytics and visualisation to facilitate the understanding of the impact of the outbreak and corresponding policies such as social distancing on transportation systems.

In general, ontology-based systems that utilise knowledge graphs and linked data are being used for data integration,

information retrieval, recommender systems as well as explainable machine learning [20]. Such approaches have the feasibility to be adopted in solving analytics challenges associated with pandemic [21]–[23]. Currently, ontology-based solutions have been largely deployed for health information exchange and communication only [24].

Given the significance of applications that can use ontological reasoning for COVID-19 data, and the lack of systematic studies that explore how ontologies are utilised in designing and supporting COVID-19 analytics, this paper aims to investigate existing solutions based on published literature on ontology-based solutions and tool support for COVID-19.

## III. METHODOLOGY

We followed the methodology of Systematic Mapping Studies (SMS) [25] to conduct and document this review. SMS is used to provide an overview (systematic map) of a research topic by showing the type of research and the results that have been published by categorising them in line with answering a specific research question [25], [26].

### A. Specifying the Research Questions (RQs)

This review is conducted to answer three research questions which are stated as follows:

- RQ1 - How ontologies are used to address the challenges related to COVID-19 analytics?
- RQ2 - What level of tool support is provided in identified studies for ontology models and applications that support COVID-19 analytics?
- RQ3 - What are the recurring challenges and reference architecture modules that emerge from investigated solutions?

### B. Identifying Solutions and Collecting Data

We followed the process proposed by Petersen et al. [25] and conducted the initial evidence search on three databases (IEEE Xplore, ACM Digital Library, and ScienceDirect) for published research on ontologies for COVID-19.

The intent behind the selection for three primary sources was that we wanted to analyse COVOD-19 specific ontologies from computing or information system point of view [25], [27]; explicitly highlighting reference architectures and tools while discarding ontology solutions that overlook the design (i.e., architecture) and implementation (tool support) of the solutions. Google Scholar search engine was used to search for literature that may have been missed by the three main sources and also to ensure the latest studies that may be available as preprints are also included as our evidence. Findings were further extended through snowballing approach proposed by Wohlin [28].

All the solutions identified from the search phase were reviewed for relevancy. If a paper satisfied the selection criteria, we included it in the list of studies qualified for the synthesis. Below are the exclusion criteria we adapted from Khan et al. [27]:

- Books and news articles
- Papers where ontologies were not incorporated to propose COVID-19 related challenges
- Vision papers
- Papers not written in English
- Full text that was not available for public access or through digital library services

Through the initial database search, we identified 55 empirical studies as candidates for review. Among those, 8 studies were qualitatively selected as relevant studies, based on the study quality assessment and exclusion criteria. The same steps were applied to the 8 studies identified through snowballing and we identified 1 additional relevant paper for our study. One additional study was included based on expert recommendations. To avoid the inclusion of duplicate studies which would inevitably bias the result of the synthesis, we thoroughly checked if very similar studies were published in more than one paper. Out of 56, we eventually selected a total of 10 studies (i.e., 18% approx. of total identified literature) that were included in the synthesis of evidence. The following section presents the results based on thematic mapping of identified studies and how they answer each of the research questions of our study.

## IV. RESULTS

To understand the nature of COVID-19 analytics applications that exploit ontology-based solutions, we mapped the ten identified studies into key themes as illustrated in Fig. 1. We identified five themes, with the majority of studies using ontologies for health monitoring, disease modelling, drug modelling, and data intelligence. One study [S1][1] uniquely proposed to utilise an ontology for telehealth applications to provide semi-automated recommendations and suggestions for telehealth patients and practitioners.

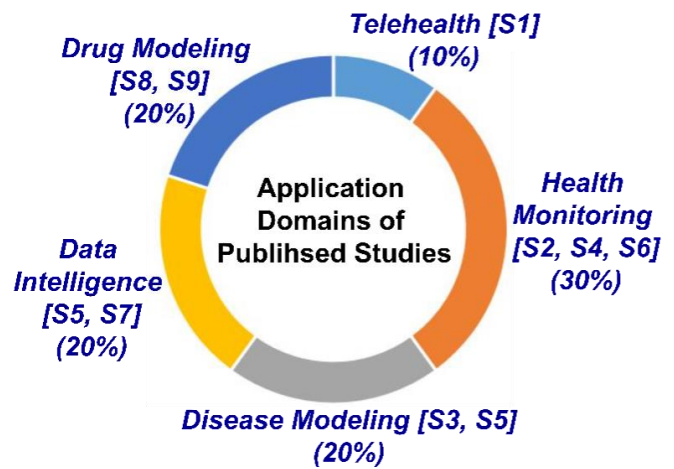

Fig. 1. Application themes of selected studies.

---

[1]The notation [SN], N is a number from 1 to 10 that represents a unique identify for each study that had been included in the review as listed in Appendix A.

## A. RQ1 - How ontologies are used to address the challenges related to COVID-19 analytics?

To answer the RQ1, we mapped the identified studies into challenges they addressed, solutions they propose, along with the method of evaluation and the ontologies they use or propose. This mapping is summarised in Table I.

Five studies [S2, S3, S4, S5, S10] propose new ontologies, and four studies [S6, S7, S8, S9] propose solutions based on existing ontologies. S1 is the only study that is not specific to a particular ontology, and proposes a placeholder to plug-in any existing ontology that suits the end-user.

## B. RQ2 - What level of tool support is provided in identified studies for ontology models and applications that support COVID-19 analytics?

To answer RQ2, we identified and characterised the tools proposed or used in identified studies, including the type of tool, their intent, source type and level of automation (Table II).

We observed that [S1, S2] proposed new tools. One study [S4] does not report on the tool(s) they used to develop the ontology they propose. The rest of the studies utilise existing open source tools and libraries such as Protégé, and KGX. Other than Protégé, identified studies tend to use a unique set of tools and libraries to match their application intent.

The level of automation of existing tools are largely observed to be limited to rule execution and data annotation. [S1, S2] are the only applications that go beyond the traditional scope and used ontologies to automate the analytics and recommendation aspects of COVID-19 information systems.

## C. RQ3- What are the recurring challenges and reference architecture modules that emerge from investigated solutions?

To answer RQ3, we created a template-based specification (serving as a structured catalogue) that can map existing studies to recurring challenges, proposed solutions and reference architectures that emerge from proposed solutions. Recurring challenges refer to challenging scenarios of COVID-19 analytics common to different application domains that need unique architecture-centric solutions to address the problems. Furthermore, the mapping we developed contains a thumbnail architectural view for the solution and examples from identified studies. This information is captured using the following topical structure:

- Recurring Challenge
- Proposed Solution
- Architectural View
- Illustrative Example

Derived from the reviewed studies, we defined four recurring challenges referred to as 1. Agent and intelligent systems; 2. Infection monitoring and analysis; 3. Integration and interoperability of disease data; and 4. Establishing information repositories. Details of the findings for each category are presented here.

### 1) Agents and Intelligent Systems:

- **Recurring Challenge:** How to exploit data-driven ontology-based intelligence for semi-autonomous systems that can assist in COVID-19 scenarios in the context of telehealth care?
- **Proposed Solution:** An ontology-based domain model structures a knowledge base that represents information about the disease, its typical symptoms, possible courses of action, and aspects to be monitored. Moreover, this domain model can structure data about the patient's characteristics and health status. A semi-autonomous telehealth system can then:
    – Reason with knowledge base as to infer the best course of action to be recommended for that particular patient
    – With the approval and monitoring of that patient's physician, activate a telehealth agent for pursuing a number of assisting actions (e.g., monitoring and transmitting specific signs of the patient, notifying the proper health authorities).
- **Architectural View:** Fig. 2

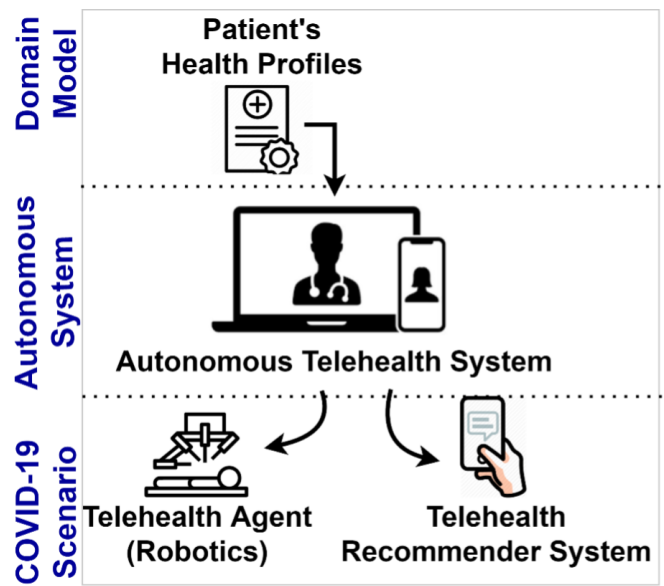

Fig. 2. Reference architectural view for ontology-based agents and intelligent systems

- **Examples:**
    S1 – Telehealth robotics
    S4 – Drug modeling
    S6 – Recommender systems

### 2) Infection Monitoring and Analysis:

- **Recurring Challenge:** How to exploit crowd-sensed data for monitoring and analysis of COVID-19 infections?
- **Proposed Solution:** A four layered architecture to support monitoring and analysis of data generated by general public. The layers are:

TABLE I
HOW ONTOLOGIES ARE USED TO ADDRESS THE CHALLENGES IN EACH STUDY

| Study ID | Challenge | Proposed Solution | Method of Evaluation | Ontology |
|---|---|---|---|---|
| [S1] | How to support robotic telehealth system for COVID-19 patient treatment? | Design and develop robotics healthcare agents where an ontology can provides knowledge representation for operations of the telehealth robotics | Prototype, Controlled Experiments | N/A[a] |
| [S2] | How to support COVID-19 surveillance for potential infections? | Ontology-based modelling and analytics of COVID-19 cases with dashboard visualisation | Use Cases, Healthcare Scenario | COVID-19 surveillance ontology |
| [S3] | How to support integration and interoperability for data from various domains? | A methodology for building novel, powerful, pathogen-specific ontologies that represent data about novel diseases that can be used to easily compare multiple dimensions of data pre-curated from past diseases. | Case Study | Coronavirus Infectious Disease Ontology |
| [S4] | How to capture data and knowledge about infectious disease? | Capturing and model disease knowledge for sharing and analysis | N/A | Coronavirus Infectious Disease Ontology |
| [S5] | How to integrate open data related to COVID-19 from various sources for decision? | Ontology-based method for data acquisition, representation, and transformation to RDF, that enables interlinking data with other publicly available data sources. | Prototype, Case study | Covid-19 ontology |
| [S6] | How to fuse data from multiple sources for COVID-19 analysis? | Novel approach that combines between the Semantic Web Services (SWS) and big data to extract information from multiple data sources to generate real-time statistics and reports | Case Study | Multiple Global and local ontologies |
| [S7] | How to develop a catalogue of solution to model COVID-19 datasets? | Semantically annotate COVID-19 Open Research Dataset | Annotation summary statistics | Existing ontologies and vocabularies |
| [S8] | How to model drugs and their impacts graph for COVID-19 medicine? | Systematic collection, annotation, and analysis of various anti-coronavirus drugs from the biomedical literature. | Prototype and controlled experiments | Existing ontologies |
| [S9] | How to identify and manage data about COVID-19 medication? | A knowledge base and a web portal to integrate an existing biomedical knowledge graph with information gathered from recently published biomedical information regarding COVID-19. | Case Study | Existing Ontology |
| [S10] | How to support text mining and semantic interoperability of unstructured data in the COVID-19 domain? | A novel ontology that forms the bases for establishing reference namespace for a COVID-19 knowledge graph that encapsulates COVID-19 domain specific topics ranging from epidemiology, and prevention and control to genetics and molecular processes. | Use Cases | The COVID-19 Ontology |

[a]Provides a placeholder to plug-in any ontology

TABLE II
AN OVERVIEW OF TOOLS THAT UTILISE ONTOLOGIES FOR COVID-19 CHALLENGES

| Study | Tool | Tool Type | Intent | Source Type | Level of Automation |
|---|---|---|---|---|---|
| [S1] | Virtual Assistant Agent | Novel Web Platform | Agent to help telehealth patients | Proprietary | Semi-automated recommendations and feedback |
| [S2] | COVID19 dashboard | Novel Mobile and Desktop application | Visualise the status of COVID-19 infections | Proprietary | Automated analytics and visualisation with user specific parameters |
| [S3] | Protégé | Standalone platform | Model and integrate data | Open Source | Rule execution |
| [S5] | RDF-Gen | Java library | Support data integration and consolidation | Open Source | Data extraction, annotation and integration |
| [S5] | Web platform | Web Framework | Querying | Open Source | N/A |
| [S6] | Protégé | Standalone Platform | Model ontology | Open Source | Rule execution |
| [S7] | TERMite | Standalone Platform | Named entity recognition tool | Proprietary | Data annotation |
| [S8] | Ontofox | Web application | Ontology extraction | Open Source | Rule execution |
| [S8] | Protégé | Standalone Platform | Ontology editing and analysis | Open Source | Rule execution |
| [S9] | KGX | Python library | Knowledge graph integration tool | Open Source | Rule execution |
| [S9] | Data Translator Node Normalisation API | Python library | Data Integration | Open Source | Rule execution |
| [S9] | Neo4j | Graph database | Graph query | Open Source | Rule execution |
| [S10] | Protégé | Standalone Platform | Ontology editing | Open Source | Rule execution |
| [S10] | Ontofox | Web application | Ontology extraction | Open Source | Rule execution |

- Sensing Layer: Contains data sensing devices (such as mobile phones, temperature sensors) that sense data regarding concepts as defined in the ontology.
- Domain Model: Stores integrated information from sensing layer, following the ontology structure.
- Analytics Layer: Contains health analytics applica-

tions that conduct analysis on the data layer.
- Presentation Layer: The analysed data is presented as dashboard and reports (e.g. number of potential infections, infections per unit area).

- **Architectural View:** Fig. 3

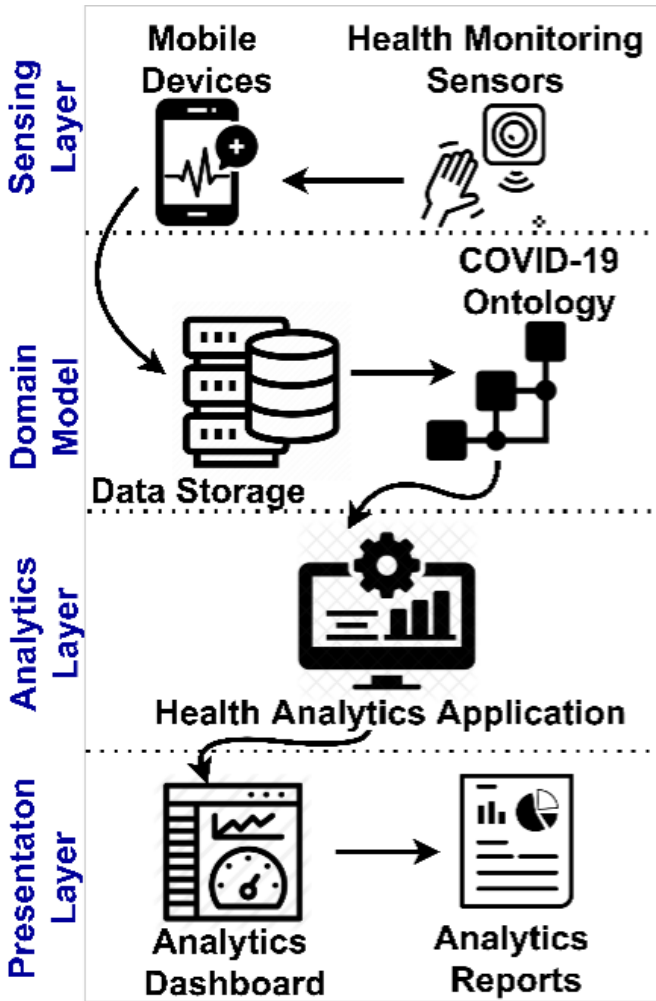

Fig. 3. Reference architectural view for infection monitoring and analysis

- **Example:**

  S2 – Pandemic Surveillance

3) *Integration and Interoperability of Disease Data*:

- **Recurring Challenge:** How to collect and represent disease data in a consistent format that can be readily queried, retrieved, and reasoned with?
- **Proposed Solution:** A three-layer architecture :
    - Sniffing Layer: Sense and extract data from different data sources manually or automatically.
    - Domain Model: Pre-process and store data following the ontology model.
    - Retrieval Layer: A uniform interface to query and view health data.
- **Architectural View:** Fig. 4
- **Example:**

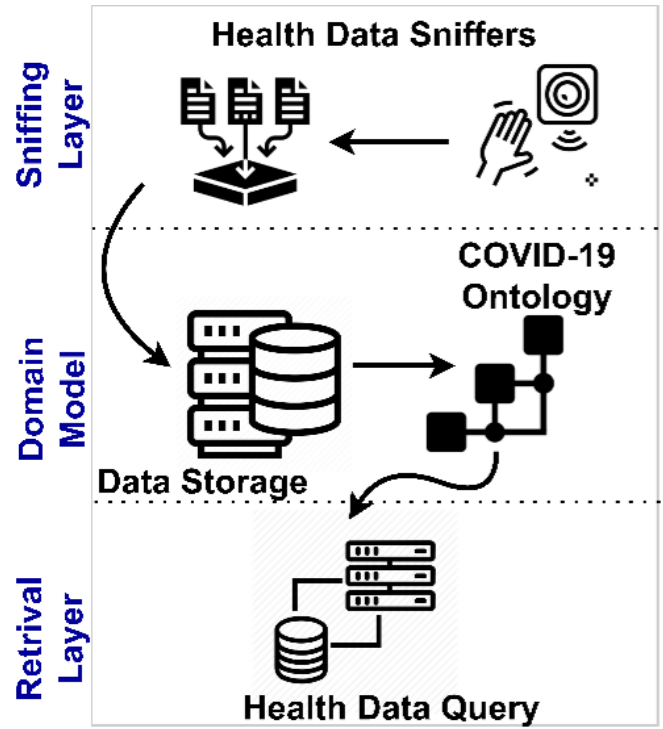

Fig. 4. Reference architectural view for integration and interoperability of disease data

  S3 – Establishing Pandemic Ontology
  S5 – Integrating Disease Data
  S9 – Integrating Disease Data

4) *Establishing Information Repositories*:

- **Recurring Challenge:** How to establish open datasets and public repositories that enable querying data for COVID-19 drugs and treatments?
- **Proposed Solution:** A three-layer architecture:
    - Public Data Layer: Represents existing public data repositories and associated APIs for open research data and biomedical literature.
    - Ontology and Data Annotation Layer: Gather data from the public data layer and annotate them via ontologies and vocabularies, to enable data modelling and integration.
    - Data Storage and Retrieval Layer: Store annotated and integrated data and provide a uniform interface to query and view data.
- **Architectural View:** Fig. 5
- **Example:**

  S7 – Annotate COVID-19 Open Research Datasets
  S8 – Annotate anti-coronavirus drugs from biomedical literature
  S10 - Annotate literature, publicly available knowledge graphs and disease maps and store data as knowledge graphs

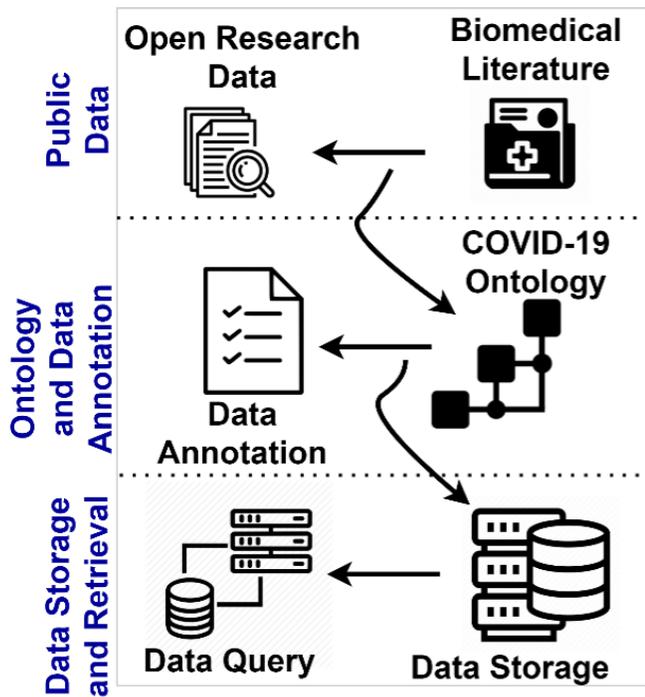

Fig. 5. Reference architectural view for information repositories

## V. Discussion

Based on the evidence gathered through three research questions under the SMS, we identified a unique set of challenges that are addressed by ontologies, related to COVID-19 analytics. We identified the following areas as emerging research trends in line with the interests of researchers and capabilities developed to date:

- Telehealth agents
- Public and open data and models for disease modelling
- Explainable statistical algorithms and data mining techniques used in infection modelling
- Drug and impact modelling, particularly for drug repurposing

Furthermore, Internet of Things (IoTs) as a network of interconnected devices and an enabling technology has exploited ontologies to represent COVID-19 specific knowledge for applications such as orchestrating sensors for context-sensing of infection transmission [15] and health symptoms monitoring [29]. Ontology-driven IoTs to manage COVID-19 represent a diverse topic that requires future work to investigate the integration of ontologies that provide a knowledge base and decision support for IoT sensors [29], [30].

A noticeable limitation in the identified studies is the lack of rigor in the evaluation of the proposed solutions. Many studies lack concrete implementations, real-world validations, and use- cases that can validate the practical applicability of proposed solutions. This issue is understandable, given the fact that COVID-19 analytics is an emerging field and its literature is immature.

## VI. Conclusion

In this paper, we provided a review of existing ontology-based solutions concerning COVID-19 analytics. We investigated challenges that have been already addressed in existing research along with tool support availability. Furthermore, we derived reference architectures from the identified studies to represent recurring challenges associated with COVID-19 analytics.

As future work, it is necessary to conduct a qualitative analysis of novel ontologies proposed in the identified studies based on existing ontology quality evaluation literature conducted on popular knowledge bases such as Wikidata and utilising formal criterion set for ontology quality evaluation [31].

## Appendix A- Selected Studies for Review

1) F. Lanza, V. Seidita, and A. Chella, "Agents and robots for collaborating and supporting physicians in healthcare scenarios," Journal of biomedical informatics, vol. 108, p. 103483, 2020.
2) S. de Lusignan, H. Liyanage, D. McGagh, B. D. Jani, J. Bauwens, R. Byford, D. Evans, T. Fahey, T. Greenhalgh, N. Jones et al., "In-pandemic development of an application ontology for covid-19 surveillance in a primary care sentinel network," JMIR Public Health and Surveillance, 2020.
3) S. Babcock, J. Beverley, L. G. Cowell, and B. Smith, "The infectious disease ontology in the age of covid-19," Journal of Biomedical Semantics, vol. 12, no. 1, pp. 1–20, 2021.
4) Y. He, H. Yu, E. Ong, Y. Wang, Y. Liu, A. Huffman, H.-h. Huang, J. Beverley, J. Hur, X. Yang et al., "Cido, a community-based ontology for coronavirus disease knowledge and data integration, sharing, and analysis," Scientific data, vol. 7, no. 1, pp. 1–5, 2020.
5) G. M. Santipantakis, G. A. Vouros, and C. Doulkeridis, "Towards integrated and open covid-19 data," arXiv preprint arXiv:2008.04045, 2020.
6) J. Kachaoui, J. Larioui, and A. Belangour, "Towards an ontology proposal model in data lake for real-time covid-19 cases prevention," 2020.
7) O. Giles, R. Huntley, A. Karlsson, J. Lomax, and J. Malone, "Reference ontology and database annotation of the covid-19 open research dataset (cord-19)," bioRxiv, 2020.
8) Y. Liu, W. Chan, Z. Wang, J. Hur, J. Xie, H. Yu, and Y. He, "Ontological and bioinformatic analysis of anti-coronavirus drugs and their implication for drug repurposing against covid-19," 2020.
9) D. Korn, T. Bobrowski, M. Li, Y. Kebede, P. Wang, P. Owen, G. Vaidya, E. Muratov, R. Chirkova, C. Bizon et al., "Covid-kop: integrating emerging covid-19 data with the robokop database," Bioinformatics, vol. 37, no. 4, pp. 586–587, 2021.
10) A. Sargsyan, A. T. Kodamullil, S. Baksi, J. Darms, S. Madan, S Gebel et al., "The COVID-19 Ontology", Bioinformatics, vol. 36, no. 24, pp. 5703–5705, 2020.